\documentclass[prd,showpacs,preprintnumbers,amsmath,amssymb,%
               notitlepage,nofootinbib]{revtex4-1}
\usepackage{graphicx}
\usepackage{color}

\newcommand{\Nf}{N_{\text{f}}}
\newcommand{\Nc}{N_{\text{c}}}
\newcommand{\MB}{M_{\text{B}}}
\newcommand{\Mpi}{M_{\pi}}
\newcommand{\mpi}{m_\pi}
\newcommand{\calC}{\mathcal{C}}

\newcommand{\calO}{\mathcal{O}}
\newcommand{\LQCD}{\Lambda_{\text{QCD}}}
\newcommand{\SU}{\mathrm{SU}}
\newcommand{\muI}{\mu_{\text{I}}}
\newcommand{\muB}{\mu_{\text{B}}}

\newcommand{\muq}{\mu_{\text{q}}}
\newcommand{\mq}{m_{\text{q}}}
\newcommand{\Mq}{M_{\text{q}}}
\newcommand{\mpiquench}{m_\pi^{\text{quench}}}
\newcommand{\feyn}[1]{
  \setbox0=\hbox{\ensuremath{#1}}
  \hbox to\wd0{\hbox to0pt{\hbox to\wd0{\hss/\hss}\hss}\box0}}

\begin{document}

\title{Silver Blaze Puzzle in $1/\Nc$ Expansions of Cold and Dense QCD Matter}

\author{Adi Armoni}
\affiliation{Department of Physics, Swansea University,
             Singleton Park, Swansea, SA2 8PP, UK}
\affiliation{Kavli Institute for the Physics and Mathematics of
             the Universe, The University of Tokyo,
             Chiba 277-8568, Japan}

\author{Kenji Fukushima}
\affiliation{Department of Physics, The University of Tokyo,
             7-3-1- Hongo, Bunkyo-ku, Tokyo 113-0033, Japan}

\begin{abstract}
  We consider quantum chromodynamics (QCD) with $\Nc$ colors and $\Nf$
  quark flavors at finite quark chemical potential $\muq$ or isospin
  chemical potential $\muI$.  We specifically address the nature of the
 ``Silver Blaze'' behavior in the framework of $1/\Nc$
  expansions.  Starting with the QCD partition function, we implement
  Veneziano's $\Nf/\Nc$ expansion to identify the density onset.  We
  find the baryon mass $\MB$ and the pion mass $\mpi$ appearing from
  different orders of Veneziano's expansion.  We argue that the
  confining properties are responsible for the Silver Blaze behavior
  in the region of $\mpi/2 < \muq < \MB/\Nc$.  We point out, however,
  that Veneziano's expansion brings about a subtlety along the same
  line as the baryon problem in finite-density quenched lattice
  simulations.  We emphasize that the large-$\Nc$ limit can allow for
  the physical ordering of $\MB$ and $\mpi$ thanks to the similarity
  between the large-$\Nc$ limit and the quenched approximation, while
  unphysical ghost quarks contaminate the baryon sector if $\Nc$ is
  finite.  We also discuss the ``orientifold'' large-$\Nc$ limit that
  does not quench quark loops.
\end{abstract}
\maketitle

\section{Introduction}

Understanding the phase diagram of quantum chromodynamics (QCD) is
one of the most pressing problems.  It is theoretically expected that
the phase diagram has rich structures as a function of the temperature
$T$, the baryon or quark chemical potential $\muB=\Nc\muq$, the
isospin chemical potential $\muI$, the external electromagnetic
fields, and so on.  Hot and/or dense QCD phases such as a quark-gluon
plasma, color superconductivity, and the hadronic phase where we live
are supposed to be realized in the Early Universe, inside of compact
stellar objects, and in the relativistic heavy-ion collisions (see
\cite{Fukushima:2010bq} for recent reviews).  The theoretical
understanding of a limited portion of the phase diagram in the
high-$T$ and low-$\muq$ region has been obtained by means of several
techniques including lattice-QCD simulations, effective model studies,
and re-summed perturbation theories.  In contrast, for the dense
system at small $T$, the understanding is far from satisfactory except
for normal nuclear matter where the traditional nuclear theory can be
applied, and for such high density that perturbative QCD works once
all gluons get screened by the Meissner effect in color
superconductivity~\cite{Rischke:2000cn,Alford:2007xm}.  Unlike the
case with $\muq\ll T$, lattice-QCD simulations based on the
Monte-Carlo algorithm has a serious difficulty called the sign problem
due to a fluctuating phase of the fermion (Dirac) determinant at
$\muq\neq0$~\cite{review:signproblem}.  There are several approaches
to evade the sign problem;  the Taylor expansion~\cite{Taylor}, the
re-weighing method~\cite{reweighting}, the imaginary chemical
potential~\cite{imaginary}, and the complex Langevin
equation~\cite{ComplexLangevin}, but they are not yet well-developed
to go beyond the small-$\muq/T$ regime.

In the $T=0$ limit, as long as $\muB$ is smaller than the baryon mass
minus the nuclear binding energy (i.e.\
$\MB-B\simeq 923\;\text{MeV}$), we can make a firm conclusion even
without tackling the sign problem.  Trivially, no physical excitation
is allowed for $\muB<\MB-B$ and none of physical properties should
depend on $\muB$ or $\muq$ then.  It is also the case for a system
with a finite isospin chemical potential $\muI$ if it is smaller than
$\mpi/2$.  Even though physics is transparent on the intuitive level,
the microscopic origin of $\muq$-independence or $\muI$-independence
has a puzzling character and still deserves theoretical
investigations, as first pointed out in Ref.~\cite{Cohen:2003kd}.
Even with sufficiently small $\muq$ or $\muI$, since the Dirac
operator and thus its eigenvalues have explicit dependence on chemical
potentials, one would naively expect that the partition function
depends on such $\muq$ or $\muI$, but physically it should not.
Interestingly, thus, that nothing happens trivially is a hint of
non-trivial physics inherent in the sign problem.  This problem is
often called the ``Silver Blaze'' named after a famous detective story
of Sherlock Holmes~\cite{Cohen:2003kd}.

In the case of $\muI$, it has been convincingly shown already in
Ref.~\cite{Cohen:2003kd} that the Dirac determinant can be
$\muI$-independent due to a gap in the energy spectrum and this gap is
given precisely by $\mpi/2$.  The same argument can hold for the
$\muq$-independence as long as $\muq \leq \mpi/2$.  Thus, the Silver
Blaze problem remains profound for the specific window,
$\mpi/2 < \muq < (\MB-B)/\Nc$.  It is argued that the phase
fluctuation should be responsible for the $\muq$-independence in this
region.  In fact, there is a demonstration that the average over the
phase fluctuations of the Dirac determinant may cancel the
$\muq$-dependence~\cite{Ipsen:2012ug}.  Such a physical mechanism with
fluctuating phase ought to be related to quark
confinement~\cite{Bringoltz:2010iy};  indeed, quark excitation is
averaged out by the $\mathrm{Z}_{\Nc}$-symmetric phase distribution in
the confined phase that is to be identified as the disordered
state~\cite{Svetitsky:1982gs}.

The main purpose of this paper is to address the Silver Blaze behavior
within the framework of Veneziano's $\Nf/\Nc$ expansion and also
't~Hooft's $1/\Nc$ expansion.  To this end we exploit the worldline
formalism (or the canonical ensemble representation) in order to
expand the Dirac determinant in powers of
$\Nf/\Nc$~\cite{Armoni:2012jw} (see also arguments in
Ref.~\cite{Bringoltz:2010iy}).  The leading order
$\calO(\Nc^2)$-contribution to the free energy is purely gluonic.  We
will focus on the sub-leading $\calO(\Nc\Nf)$- and the sub-sub-leading
$\calO(\Nf^2)$-contributions.  At $\calO(\Nc\Nf)$ order two theories,
one with $\muI\neq0$ and $\muq=0$ and the other with $\muq\neq0$ and
$\muI=0$ are equivalent to each other in the $\calC$-even
sector~\cite{Armoni:2012jw,Cohen:2004mw,Hanada:2011ju}, as long as
$\muI$ and $\muq$ are small enough not to induce any finite baryon or
isospin density.  When the next $\calO(\Nf^2)$-contribution is added,
the equivalence between $\muI$ and $\muq$ no longer
holds~\cite{Cohen:2004mw}.  The virtue of this $\Nf/\Nc$ expansion
lies in the clear separation of the baryon and pion sectors that
belong to the $\calO(\Nc\Nf)$- and the $\calO(\Nf^2)$-contributions,
respectively.

Although our formulation provides us with a useful organization of
different physics origins, Veneziano's expansion at finite density is
a subtle expansion and we should be cautious about the results.  In
fact, the $\Nf/\Nc$ expansion amounts to a series of
$\Nf/\Nc$ corrections around the state at $\Nf/\Nc\to0$ that is
nothing but the quenched limit.  It is known that the quenched
simulation at finite density \footnote{This may sound peculiar, as the
  Yang-Mills theory is density free.  Here, finite-density quenched
  simulations refer to taking the expectation value of finite-density
  operators, which are typically non-Hermitian, with the vacuum of the
  Yang-Mills theory.} sometimes leads to unphysical results,
especially the density onset seems to be set not by $\MB$ but $\mpi$,
which would cause additional subtlety in the argument on the Silver
Blaze puzzle in Veneziano's expansion.  We argue that in the 't~Hooft
limit of $\Nc\to\infty$ the formulation could be put in a rather clean
environment and our argument is validated.  With finite $\Nc$ or in
the orientifold large-$\Nc$ limit as we discuss later, it seems that
the Silver Blaze puzzle still remains quite non-trivial.

Our paper is organized as follows:  In Sec.~\ref{sec:formalism}, we
review the decomposition of the QCD partition function in the
$\Nf/\Nc$ expansion with a chemical potential.  In
Sec.~\ref{sec:NcNf}, with $\Nf=2$ fixed, we analyze the expression for
the free energy at $\calO(\Nc\Nf)$ and discuss the dependence of the
free energy on $\muI$ and $\muq$.  Next, in Sec.~\ref{sec:Nf2}, we
turn to the next-order free energy at $\calO(\Nf^2)$.  We briefly discuss
 a possible situation in the orientifold large-$\Nc$ limit
in Sec.~\ref{sec:orientifold}.  Finally, Sec.~\ref{sec:Discussion} is
devoted to discussions and conclusions.

\section{Veneziano Expansion of the Free Energy}
\label{sec:formalism}

\begin{figure}
 \includegraphics[width=0.2\textwidth]{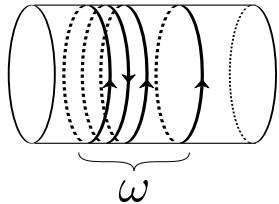}
 \caption{Schematic picture of the decomposition of the Dirac
   determinant according to the winding number $\omega$ that
   corresponds to how many times the configuration wraps around
   $S^1$.  This example shows a case of $\omega=3-1=2$.}
 \label{fig:schematic}
\end{figure}

Let us consider QCD generalized for $\Nc$ colors and degenerate $\Nf$
quark flavors with identical mass $\mq$.  The partition function of
this theory on $R^3\times S^1$ with the anti-periodic boundary
condition for quarks is expressed in a form of the functional
integration with respect to the gauge fields $A_\mu$.  The integrand
consists of the weight factors from the Yang-Mills action (denoted by
$\exp[-S_{\rm YM}(A)]$) and the fermionic Dirac determinant (denoted
by $\exp[\Gamma(\mu_f)]$).  The latter can be decomposed generally in
terms of the winding number $\omega$ as
\begin{equation}
 \Gamma(\mu_f) = \ln \det(\feyn{D} + \mq + \mu_f \gamma^0)
 = \sum_f \sum_{\omega=-\infty}^\infty \Gamma_f^{(\omega)}(\mu_f)
 = \sum_f \sum_{\omega=-\infty}^\infty \Gamma_f^{(\omega)}(0)\,
   e^{\mu_f\omega/T}\;.
\end{equation}
Here, the index $f$ runs over different quark flavors.  The explicit
expression of $\Gamma_f^{(\omega)}$ as a function of $A_\mu$ can be
found in the literature, e.g.\ with use of the worldline
formalism~\cite{Armoni:2012jw}.  We note that this decomposition with
respect to $\omega$ can translate into the canonical ensemble with
$\omega$ identified as the quark
number~\cite{Miller:1986cs,Fukushima:2002bk}.
Figure~\ref{fig:schematic} shows an example of the $\omega=2$ case.
We note that the configuration along $S^1$ may be wandering, and a
special straight configuration is nothing but the Polyakov loop.

An important property of $\Gamma$ is that the connected $k$-point
function of $\Gamma$ is diagrammatically suppressed by $\Nf/\Nc$
as~\cite{Armoni:2012jw}
\begin{equation}
 \bigl\langle \underbrace{\Gamma \Gamma \cdots \Gamma }_{k \text{ times}}
 \bigr\rangle_{{\rm c;YM}} \sim \Nc^2\Bigl({\Nf\over \Nc}\Bigr)^{k} \;.
\label{eq:veneziano}
\end{equation}
This $\Nc^{-k}$ suppression appears from the gluon interaction
$g^2\sim \calO(\Nc^{-1})$ that connects $\Gamma$'s.  The expansion of
$\langle\exp(\Gamma)\rangle$ in terms of $\Gamma$ (after taking the
flavor sum) leads to an expansion in powers of $\Nf/\Nc$, namely, the
expansion in the Veneziano limit.  Here, we take the average in the
vacuum at $\Nf/\Nc=0$, i.e.\ the pure Yang-Mills theory whose weight
factor is $\exp({-S_{\rm YM}})$.  The subscript ``c'' in the
expectation value denotes the contribution from the ``connected''
diagrams.  Another important property of $\Gamma$ is that it is
exponentially suppressed with $\omega$ at least as
\begin{equation}
 \Gamma_f^{(\omega)}(0) \;\sim\;
  \exp \Bigl(-{\mq \over T}|\omega| \Bigr)
\label{eq:sup_wind}
\end{equation}
with the bare quark mass $\mq$~\cite{Armoni:2012jw}.  The suppression
could be even faster with $\mq$ replaced with a dynamical $\Mq$ when
we take the expectation value, $\langle\Gamma_f^{(\omega)}(0)\rangle$,
which we will discuss later.  This property ensures the convergence of
the expansion in terms of $\omega$ up to a certain value of $\mu_f$
less than the quark mass.

The free energy is then expressed in a form of an expectation value as
\begin{equation}
 F = F_{\rm YM}  -T\ln\bigl\langle e^{\Gamma(\mu_f)} \bigr\rangle_{\rm YM}
   = F_{\rm YM} + F^{(1)} + F^{(2)} + \calO(\Nc^{-1}\Nf^3)
\label{expansion}
\end{equation}
with
\begin{align}
 F_{\rm YM} & ~\sim~ \calO(\Nc^2) \;,\\
 F^{(1)} & = -T \sum_f \sum_{\omega=-\infty}^\infty
  \bigl\langle \Gamma_f^{(\omega)}(0) \bigr\rangle_{\rm YM}\;
  e^{\mu_f\omega/T} ~\sim~ \calO(\Nc\Nf) \;,
\label{eq:f1}\\
 F^{(2)} & = -T \sum_{f,f'}\sum_{\omega,\omega'=-\infty}^\infty
  \bigl\langle \Gamma_f^{(\omega)}(0)\Gamma_{f'}^{(\omega')}(0)
  \bigr\rangle_{\rm c;YM}\; e^{(\mu_f\omega+\mu_{f'}\omega')/T}
  ~\sim~ \calO(\Nf^2) \;,
\label{eq:f2}
\end{align}
where $F_{\rm YM}$ represents the pure gluonic energy of
$\calO(\Nc^2)$, the second $F^{(1)}$ with
$\langle \Gamma_f^{(\omega)}(0)\rangle_{\rm YM}$ (in which we omitted
``c'' that is irrelevant for the one-point function of
$\Gamma_f^{(\omega)}(0)$) is of $\calO(\Nc\Nf)$, and the expansion
goes on as the third $F^{(2)}$ with $\langle
\Gamma_f^{(\omega)}(0)\Gamma_{f'}^{(\omega')}(0) \rangle_{\rm c;YM}$
of $\calO(\Nf^2)$ and so on, according to Eq.~\eqref{eq:veneziano}.
It is important to note that this is not yet a consistent ordering of
the $1/\Nc$ expansion;  our identification of $F^{(1)}$ and $F^{(2)}$
is based on the power of $\Nf/\Nc$ and each of $F^{(1)}$ and $F^{(2)}$
contains sub-leading (non-planar) contributions suppressed by higher
powers of $1/\Nc$.

In this paper we shall work only at sufficiently small temperature,
$T\ll \LQCD$, where the Yang-Mills vacuum should be in the confined
phase.  (Otherwise, it is rather trivial what is going on in the
deconfined phase.)  This phase is characterized by the realization of
$Z_{\Nc}$ (center) symmetry, so that the expectation value of a
center non-symmetric operator vanishes, namely,
\begin{equation}
 \bigl\langle \Gamma ^{(\omega)} \bigr\rangle_{\rm c;YM} =
  \left\{ \begin{array}{cp{1em}l}
   \text{non-zero} && \text{for $\omega=0\;{\rm mod}\;\Nc$} \\
   0 && \text{otherwise}
  \end{array} \right. \;.
\end{equation}
It should be mentioned that the expansion is made around the vacuum of
the pure Yang-Mills theory, so that confinement can bear a
well-defined meaning and the above expectation value can be strictly
vanishing except for $\omega=0, \Nc, 2\Nc, \dots$.  Physically
speaking, one winding corresponds to a single-quark excitation, see
Fig.~\ref{fig:schematic}, and each time the winding number reaches
$\Nc$, a color singlet is formed out of $\Nc$ quarks.  We shall thus
call such a configuration with $\omega=\Nc$ a
\textit{baryonic configuration}.  We can easily extend the above to
more general correlations in the confined phase as
\begin{equation}
 \bigl\langle \Gamma^{(\omega_1)} \Gamma^{(\omega_2)} \cdots
  \Gamma^{(\omega_k)} \bigr\rangle_{\rm c;YM} =
  \left\{ \begin{array}{cp{1em}l}
   \text{non-zero} && \text{for $\omega_1+\omega_2+\cdots+\omega_k=
   0\;{\rm mod}\;\Nc$} \\
   0 && \text{otherwise}
  \end{array} \right. \;.
\end{equation}
Then, in the confined phase at low $T$, non-vanishing terms out of
Eqs.~\eqref{eq:f1} and \eqref{eq:f2} turn out to be
\begin{align}
 F^{(1)} &= -T\sum_f \sum_{\bar{\omega}=-\infty}^\infty
  \bigl\langle\Gamma_f^{(\bar{\omega}\Nc )}(0)\bigr\rangle_{\rm YM}\;
  e^{\bar{\omega}\Nc\mu_f /T} \;,
\label{eq:f1conf}\\
 F^{(2)} & = -T\sum_{f,f'}\sum_{n=-\infty}^\infty \biggl\{
  \bigl\langle \Gamma_f^{(n)}(0)\Gamma_{f'}^{(-n)}(0)
  \bigr\rangle_{\rm c;YM}\; e^{(\mu_f-\mu_{f'})n/T} +
  \bigl\langle \Gamma_f^{(n+\Nc)}(0)\Gamma_{f'}^{(-n)}(0)
  \bigr\rangle_{\rm c;YM}\; e^{[\mu_f(n+\Nc)-\mu_{f'}n]/T}
  + \cdots \biggr\} \;,
\label{eq:f2conf}
\end{align}
where the ellipsis represents other contributions such as
$(\omega=n+2\Nc,\omega'=-n)$, $(\omega=n+3\Nc,\omega'=-n)$, and so on.

In the hadron language, intuitively, $F^{(1)}$ above corresponds to a
``multi baryon contribution'' with $\bar{\omega}$ baryons.  The first
term in the next contribution, $F^{(2)}$, corresponds to a
``mesonic contribution'' and the second in the parentheses is a mixed
correlation of baryons and mesons.  As long as $\mu_f$ is small enough
as compared to the baryonic scale of $F^{(1)}$, these mixed-type
contributions are always more suppressed than the pure mesonic
contribution.  Thus, we can safely neglect the mixed-type term when we
discuss the density region up to the onset as in what follows.

Here, let us emphasize that this procedure to take the
$\mathrm{Z}_{\Nc}$-symmetric average is a vital step to understand the
Silver Blaze problem for the baryon density onset.  As we mentioned,
one must take account of the phase fluctuations of the Dirac
determinant in this density region of $\mpi/2<\muq<\MB/\Nc$, and we
effectively do this by dropping center non-symmetric operators.
Indeed, as argued in Ref.~\cite{Fukushima:2002bk}, fractional (not a
multiple of $\Nc$) excitations of quarks that break center symmetry
explicitly are closely related to the sign problem.  Usually, in the
thermodynamic (i.e.\ infinite volume) limit in particular, the
canonical ensemble becomes quite singular and it loses the strength to
solve the sign problem practically.  In our present formulation,
however, we combine it with the $\Nf/\Nc$ expansion, so that unwanted
quark excitations diminish and the whole machinery is under
theoretical control.

\section{Large-$\Nc$ Counting of $F^{(1)}$}
\label{sec:NcNf}

In this section, we consider $F^{(1)}$ in an $\SU(\Nc)$ theory with
$\Nf=2$ fixed, and denote corresponding chemical potentials by $\mu_1$
and $\mu_2$.  In this way we can introduce a quark (baryon) chemical
potential as $\mu_1=\mu_2=\muq=\muB/\Nc$ or an isospin chemical
potential as $\mu_1=-\mu_2=\muI$.  Then, up to this order of $\Nc\Nf$,
the free energy reads
\begin{equation}
 F^{(1)}/T = -\sum_{\bar{\omega}=-\infty}^\infty
  \langle\Gamma_1^{(\bar{\omega}\Nc)}(0)\rangle_{\rm YM}\;
  e^{\bar{\omega}\Nc\mu_1/T}  -\sum_{\bar{\omega}=-\infty}^\infty
  \langle\Gamma_2^{(\bar{\omega}\Nc)}(0)\rangle_{\rm YM}\;
  e^{\bar{\omega}\Nc\mu_2/T} \;.
\end{equation}
It is crucially important to note that no difference arises at this
order for $\mu_1=\mu_2=\muq$ and $\mu_1=-\mu_2=\muI$.  In other words,
the system at finite quark chemical potential is equivalent to the
system at finite isospin chemical potential at $\calO(\Nc\Nf)$ because
there is no correlation function involving different flavor sectors.
The two flavor sectors do not talk to each other, so to
speak~\cite{Cohen:2004mw}.

Specifically in this section, we shall use a collective notation $\mu$
not distinguishing $\muq$ and $\muI$.  As we already discussed in the
previous section, in the tree-level,
$\Gamma^{(\bar{\omega}\Nc)}(0)\sim\exp(-|\bar{\omega}|\Nc\mq/T)$
implies that the free energy in the limit of $T=0$ becomes independent
of $\mu$ if $\mu<\mq$.  When we perform the functional integration over
the gauge fields, this exponential factor would decrease faster;
$|\bar{\omega}|\Nc\mq$ should be replaced with the baryonic-dressed
mass $\bar{\omega} \MB$ (where we picked up only the contribution from
$\omega$ baryons but neglected any ``composite-baryon'' possibility,
which should be empirically reasonable).  This means that
\begin{equation}
 \Gamma^{(\bar{\omega}\Nc)}(0) \;\sim\; \exp(-|\bar{\omega}|\Nc\mq/T)
 \quad\rightarrow\quad
 \bigl\langle\Gamma^{(\bar{\omega}\Nc)}(0)\bigr\rangle_{\rm YM} \;\sim\;
  \exp(-\bar{\omega} \MB/T) \;.
\label{eq:def_baryon}
\end{equation}
In other words we can state that this is our \textit{definition} of
the baryon mass.  In fact, together with Eq.~\eqref{eq:f1conf}, we can
see that the expansion takes a form of
\begin{equation}
 F^{(1)}/T \;\sim\; -\sum_{\bar{\omega}=-\infty}^\infty \exp\bigl[
\bar{\omega}(\Nc\mu - \MB )/T \bigr]
\label{eq:baryon}
\end{equation}
apart from prefactors that are not of our interest to locate the
density onset.  It is obvious that the free energy should not change
with $\mu$ until $\mu$ hits the onset at $\mu_{\rm c} = \MB/\Nc$
(which should be corrected by the binding energy $B$ of nuclear matter
that is incorporated, in principle, in the
definition~\eqref{eq:def_baryon} for large $\omega$) because no
particle can excite at $T=0$.  Thus, from this point of view of the
density onset, our definition makes sense to characterize the baryon
mass.

In the context of the Silver Blaze problem, a more non-trivial
question is whether our $\MB$ can behave differently from the pion
mass $\mpi$ or not.  The main concern regarding the Silver Blaze
problem lies in the observation that the lowest excitation energy even
in the baryonic sector seems to be governed by $\mpi$.  We therefore
need to treat our $\MB$ very carefully and should clarify if
$\MB\simeq (\Nc/2)\mpi$ or not.  If this happened unfortunately, the
results are unphysical and any useful information on the Silver Blaze
puzzle in the most non-trivial region is not available at all.  As a
matter of fact, in the realistic world with $\Nc=3$, it is often the
case that $\MB\simeq (\Nc/2)\mpi$ is concluded.

For the case with $\Nc=3$ we can have a valuable hint from 
lattice-QCD simulations.  Along the line of the lattice-QCD setup, it
would be instructive to rewrite our $F^{(1)}$ in a slightly different
form using the quark number operator $N(\mu)$.  It is easy to confirm
the following expression,
\begin{equation}
 F^{(1)} = -T\langle \Gamma(\mu)\rangle_{\rm YM}
 = -T\int^\mu d\mu'\, \biggl\langle \frac{d\Gamma(\mu')}{d\mu'}
  \biggr\rangle_{\rm YM}
 = -\Nf\int^\mu d\mu' \langle N(\mu')\rangle_{\rm YM},
\label{eq:F1density}
\end{equation}
up to a $\mu$-independent constant.  In the final form
$\langle N(\mu)\rangle_{\rm YM}$ is the same quantity as the quark
number expectation value measured in the quenched simulation.  We note
that this expectation value contains all gluonic loops, i.e., not only
planar diagrams but also higher genus diagrams, but no quark loops.
Surprisingly, the results from lattice simulations and also from
random matrix model imply that $\langle N(\mu)\rangle_{\rm YM}$
becomes non-zero when $\mu$ exceeds
$\mpi/2$~\cite{earlyonset,Gibbs:1986ut,Barbour:1986jf}.  Strictly
speaking, this $\mpi$ is not necessarily the physical pion mass, but
the quenched pion mass, $\mpiquench$.  It is still possible to
distinguish $\mpiquench$ from physical baryon mass by looking at how
they behave with decreasing $\mq$;  in the $\mq\to0$ limit $\mpi$ or
$\mpiquench$ goes to zero but the physical baryon mass should not.  If
one finds $\MB$ approaching zero in the chiral limit, $\MB$ should be
more like the pion mass rather than the physical baryon mass.

Such a striking observation was established first in the so-called
``phase quenched'' simulation, in which the fluctuating phase of the
Dirac determinant is neglected and its modulus,
$|\exp[\Gamma(\mu)]|^2$, is implemented in the simulation.  It is
understood today that such an approximation is equivalent to replacing
the chemical potential with the isospin one, $\muI$, so that the onset
is determined by not $\MB/\Nc$ but $\mpiquench/2$.  Later on, it was
recognized that the same conclusion is drawn to the quenched
simulation in which the whole Dirac determinant is neglected, which is
much more non-trivial to understand.

\begin{figure}
 \includegraphics[width=0.55\textwidth]{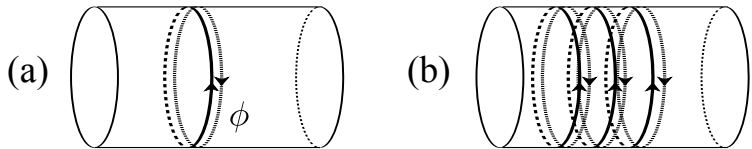}
 \caption{(a) A bound state of the baryonic pion with a combination of
 the ghost $\phi$.  (b) A baryonic configuration screened by $\phi$
 leading to three baryonic pions, the mass of which has nothing to do
 with the physical baryon mass but is characterized by
 $3m_{q\bar{\phi}}$.}
 \label{fig:schematic2}
\end{figure}

The onset at $\mpiquench/2$ in the quenched simulation is caused by
the condensation of an ``unphysical bound state'' of a quark and a
conjugate anti-quark called the baryonic
pion~\cite{Gibbs:1986ut,baryonicpion,Stephanov:1996ki} as sketched in
Fig.~\ref{fig:schematic2}~(a).  Let us explain what is happening using
the language of the so-called Partially Quenched Chiral Perturbation
Theory~\cite{PQChPT}.  In the quenched limit all quark loops should be
removed, and we can formulate this  by introducing a ghost field
$\phi$.  To cancel the Dirac determinant exactly, $\phi$ should be a
bosonic quark (but should satisfy the anti-periodic boundary condition
along the thermal $S^1$) that yields an inverse of the Dirac
determinant.  We must utilize such a formulation with quarks and
ghosts to deal correctly with the computation of non-Hermitian
expectation value like Eq.~\eqref{eq:F1density}.  Because such bosonic
ghosts are abundant at finite density, a quark can easily pick an
anti-$\phi$ up and form a bound state $q$-${\bar{\phi}}$ or the
baryonic pion, the mass of which is denoted here as $m_{q\bar{\phi}}$.
In this setup of the quenched limit, if we have a configuration with
$\omega$ quarks, as is depicted in Fig.~\ref{fig:schematic2}~(b), they
turn into $\omega$ baryonic pions rather than physical hadrons.  Thus,
we trivially have $\mpiquench=2m_{q\bar{\phi}}$, and for the baryonic
configuration with $\Nc$ quarks, $\MB$ does not access the genuine
baryonic sector but simply $\MB=\Nc m_{q\bar{\phi}}$, which
immediately leads to the funny observation,
$\MB=(\Nc/2)\mpiquench$.  In this way we can understand the subtle
nature of the quenched limit when involving non-Hermitian operators
\footnote{We can understand this also from the Dirac eigenvalues;  the
  Banks-Casher type formula for the quark number operator needs an
  eigenvalue density that can be well-defined only for a
  one-dimensional distribution.  The non-Hermiticity makes the
  eigenvalues spread over the complex plane, and to avoid this, a
  conjugate sector should be augmented.}.

From the above argument it is highly conceivable that taking the
large-$\Nc$ limit may cure the subtle situation.  The essential point
is that the large-$\Nc$ limit already encompasses the quenched limit
and no quark loops appear.  This means that we do not have to
introduce the ghost field $\phi$ to cancel the Dirac determinant.
Then, because there is no $\phi$, the theory does not have the
unphysical baryonic pion.  Of course, one can still keep introducing
$\phi$, but its excitations are negligible as compared to the gluon
excitation that is of $\calO(\Nc^2)$.  In other words, forming an
unphysical bound state is regarded as the screening effect or the
QCD-string breaking.  In the large-$\Nc$ limit the QCD string extends,
so that the linear potential and thus confinement can persist
strictly.  Because there is no $\phi$-induced screening in the
large-$\Nc$ limit, the baryonic configuration couples to the physical
baryon excitation.

To strengthen our argument, let us attempt to confirm explicitly that
$\MB/\Nc$ is certainly heavier than $m_\pi/2$ within the framework
based on the large-$\Nc$ limit.  As we already mentioned, $F^{(1)}$
contains all the sub-leading terms in the large-$\Nc$ counting, while
it is a leading-order contribution in the Veneziano expansion.  Here,
because we are interested in the behavior of $\MB$ only, it is
sufficient for us to focus on the analysis of
$\langle\Gamma^{(\Nc)}(0)\rangle_{\rm YM}$.  Then, let us pay our
close attention to the large-$\Nc$ expansion of this quantity that
consists of $\Nc$ quarks propagating in the same direction.

\begin{figure}
 \includegraphics[width=0.6\textwidth]{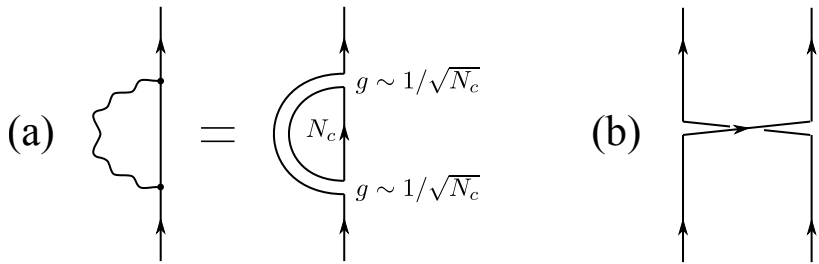}
 \caption{(a) Lowest-order diagram of the quark self-energy in the
   double-line notation.  This and all the rainbow types of
   higher-order diagrams are of the same $\Nc$ order as the
   bare-quark propagation, the sum of which should lead to the
   dynamical mass $\Mq$.  (b) One gluon exchange between two quarks,
   which is suppressed by $g^2\sim 1/\Nc$.  (There appears no $\Nc$
   factor by closing a quark loop.)}
 \label{fig:selfenergy}
\end{figure}

In each quark propagation the self-energy insertion from the
interaction with gluons may appear and this is not suppressed by
$1/\Nc$ as is illustrated in Fig.~\ref{fig:selfenergy}~(a).  We can
also think of higher-order planar diagrams of the self-energy
(typically represented by a rainbow-type re-summation), which
eventually leads to the dynamical quark mass $\Mq$ (as is the case in
the Dyson-Schwinger studies~\cite{Aoki:1990eq}).  This is how the
constituent picture of quarks with dynamical mass emerges.  For the
baryonic configuration, we can also think of different types of
diagrams that connect separate quark lines with gluons but the
interaction among quarks is always suppressed by $1/\Nc$ as explained
in Fig.~\ref{fig:selfenergy}~(b).  Each time one gluon exchange
occurs between two quarks (i.e.\ diquark interaction), it gives a
suppression factor by $g^2\sim 1/\Nc$, which is eventually compensated
for by a combinatorial enhancement.  Thus, we can parametrically write
the leading-$\Nc$ baryon mass as
\begin{equation}
 \MB \simeq \Nc \Mq - \frac{\Nc(\Nc-1)}{2}\,\cdot\,
  \frac{V_{\text{diquark}}}{\Nc} \;,
\end{equation}
where $V_{\text{diquark}}$ represents the energy gain from the
one-gluon exchange interaction in the diquark channel (scaled by $\Nc$
so that $V_{\text{diquark}}\sim\calO(1)$).  It should be also
mentioned that we neglected the kinetic energy in the above
expression, which is suppressed in the large-$\Nc$ case.  The
combinatorial factor, $\Nc(\Nc-1)/2$ originates from the number of
independent diagrams in which two of $\Nc$ quarks are selected out.
Before proceeding further, we shall look at the pion sector next.

\section{Large-$\Nc$ Counting of $F^{(2)}$}
\label{sec:Nf2}

Let us next consider the ``mesonic contribution'' of $\calO(\Nf^2)$
that makes a discrimination between $\muq$ and $\muI$.  We first consider
the isospin chemical potential, $\mu_1=-\mu_2=\muI$.  Let us remember
that, on the one hand, at $\calO(\Nc\Nf)$ the free energies with
either an isospin or a quark chemical potential are equivalent to each
other, but $F^{(2)}$ at $\calO(\Nf^2)$, on the other hand, makes a
sharp contrast and distinguishes one from the other.  The free energy
then reads,
\begin{equation}
 F^{(2)}(\muI)/T = -\sum_{n=-\infty}^\infty \bigl\langle
  \Gamma_1^{(n)}(0) \Gamma_1^{(-n)}(0)
  + \Gamma_2^{(n)}(0)\Gamma_2^{(-n)}(0) \bigr\rangle_{\rm c;YM}
  - 2\sum_{n=-\infty}^\infty \bigl\langle \Gamma_1^{(n)}(0)
  \Gamma_2^{(-n)}(0) \bigr\rangle_{\rm c;YM} \;e^{2\muI n/T} \;.
\label{eq:f2muI}
\end{equation}
In the same way as the analysis in the previous section we can define
our pion mass from the following expectation value (under the
approximation that we neglect ``composite-pion'' configurations, which
is justified in the large-$\Nc$ limit where the meson interaction is
turned off);
\begin{equation}
 \langle\Gamma_f^{(n)}(0)\Gamma_{f'}^{(-n)}(0)\rangle_{\rm c;YM}
  \;\sim\; \exp(-n \Mpi/T) \;.
\label{eq:pion}
\end{equation}
Then, using this definition of $\Mpi$, apart from unimportant
prefactors, we see that the last term of Eq.~\eqref{eq:f2muI} has the
following form of the expansion;
\begin{equation}
 \sim \sum_{n=-\infty}^\infty \exp\bigl[-n (\Mpi-2\muI)/T\bigr] \;.
\end{equation}
Thus, as long as $\muI < \Mpi/2$, the expansion is converging and the
free energy in the $T=0$ limit is completely insensitive to $\muI$,
leading to zero isospin density.  In other words the threshold of the
pion condensation is given by $\Mpi/2$, and so it is quite reasonable
to adopt the above definition~\eqref{eq:pion} of the pion mass.

For the case of the quark chemical potential, $\mu_1=\mu_2=\muq$,
unlike the isospin chemical potential case, we can see that the
leading contribution of $F^{(2)}$ is independent of $\muq$ as
\begin{equation}
 F^{(2)}(\muq)/T = -\sum_{n=-\infty}^\infty \bigl\langle
  \Gamma_1^{(n)}(0)\Gamma_1^{(-n)}(0)
 +\Gamma_2^{(n)}(0)\Gamma_2^{(-n)}(0)
 +2\Gamma_1^{(n)}(0)\Gamma_2^{(-n)}(0) \bigr\rangle_{\rm c;YM} \;.
\label{eq:mesonic-q}
\end{equation}
Therefore, the onset of the quark number density is solely determined
by $F^{(1)}$ and so our baryon mass $\MB$ gives the threshold.  It is
an interesting and non-trivial observation that the physics of $\muq$
and that of $\muI$ belong to different sectors in the power counting
of $\Nf/\Nc$.

\begin{figure}
 \includegraphics[width=0.25\textwidth]{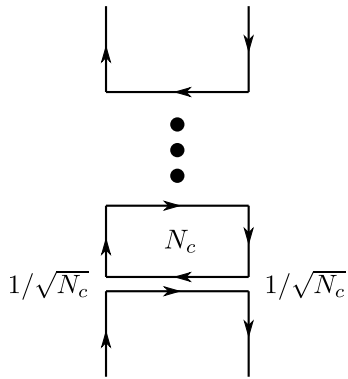}
 \caption{Ladder interactions between a quark and an anti-quark that
   forms the meson.  Lines at the top and the bottom are closed by the
   anti-periodic boundary condition and the winding number counts how
   many times the configuration wraps around this circle.}
 \label{fig:pion}
\end{figure}

Now let us proceed to a more quantitative aspect of the Silver Blaze
problem;  the critical question is how large $\Mpi/2$ is precisely?
As we argued in the previous section, the quenched simulation with
finite $\Nc$ implicitly requires the bosonic ghost fields, while quark
loops just decouple in the large-$\Nc$ limit.  The meson diagrams at
large $\Nc$ are well-known and the ladder re-summation as sketched in
Fig.~\ref{fig:pion} gives the meson.  We can then parameterize the
meson mass as follows;
\begin{equation}
 \Mpi = 2\Mq - \Nc \cdot \frac{V_{\text{scalar}}}{\Nc} \;.
\end{equation}
It is known~\cite{Buballa:2003qv} that the projection of the one-gluon
exchange interaction to the scalar and the diquark channels,
respectively, leads to the factor;
\begin{equation}
 V_{\text{diquark}} \propto \frac{\Nc+1}{2\Nc}\;,\qquad
 V_{\text{scalar}} \propto \frac{\Nc^2-1}{\Nc^2}\;,
\end{equation}
which means that
$V_{\text{diquark}}=[\Nc/2(\Nc-1)]V_{\text{scalar}}$.
For the lightest pseudo-scalar (i.e.\ pion) channel
$V_{\text{scalar}}$ should be about $2\Mq$ at most in order to realize
small $\Mpi$ of the Nambu-Goldstone boson.  Using the above
coefficients we can make an estimate of the difference between $\MB$
and $\Mpi$ as
\begin{equation}
 \frac{\MB}{\Nc} - \frac{\Mpi}{2}
  \simeq \frac{V_{\text{scalar}}}{4} > 0 \;.
\end{equation}
In particular, when the maximally large
$V_{\text{scalar}}\simeq 2\Mq$ is realized to render the pion mass to
vanish, the lower bound of the baryon mass could be
$[\MB]_{\text{lowest}} \simeq \Nc\Mq/2$, that is, the baryon mass
cannot be lighter than a half of the sum of the constituent quark
mass.  Although the quantitative estimate here might be a bit
oversimplified, the essential point in this present argument is that
it is very likely that $\MB$ is heavier than $(\Nc/2)\Mpi$.  This hand-waving argument suggests that the diquark interaction is not
strong enough to make the baryon as light as the pion in the
large-$\Nc$ world, which makes a sharp contrast to the finite-$\Nc$
quenched world where $\MB\to0$ in the chiral limit.

\section{``Orientifold'' Large-$\Nc$ Expansion}
\label{sec:orientifold}

Apart from the 't~Hooft and the Veneziano large-$\Nc$ expansions there
is another large-$\Nc$ expansion that goes under the name,
``Orientifold Expansion''~\cite{ASV}.  Consider an $\mathrm{SU}(\Nc)$
gauge theory coupled to $\Nf$ fermions that transform in the two-index
anti-symmetric representation, denoted by $\psi_{[ij]}$.  For
$\mathrm{SU}(3)$ a Dirac fermion that transforms in the anti-symmetric
representation is equivalent to a Dirac fermion that transform in the
fundamental representation, since
$q^k = \frac{1}{2}\epsilon ^{ijk}\psi_{[ij]}$.  In the large-$\Nc$
limit the fermions that carry two-indices behave like gluons.  In
particular, fermions are not quenched in the $\Nc\to\infty$ limit.
This is the most significant difference between the
``orientifold'' expansion and the 't~Hooft expansion.

The color singlets of the ``orientifold'' theory contain mesons and
baryons (in addition to glueballs).  The meson, as in ordinary QCD
consists of a pair of fermion and anti-fermion.  As for the baryons,
the issue is more subtle: the most natural candidate consists of $\Nc$
fermions, contracted by two epsilon tensors.  It turns out, however,
that this identification is not correct.  It was shown by
Bolognesi~\cite{bolognesi} (see also Ref.~\cite{Cherman:2006iy}) that
this simple ``baryon'' is not stable and, moreover, does not admit the
properties of baryons, as anticipated from the Skyrme model.  The
correct object that should be identified as the baryon consists of
$\frac{1}{2}\Nc(\Nc-1)$ fermions.

Our current discussion of the orientifold theory at finite temperature
and density is similar to the previous discussion.  We can use the
worldline formalism to expand the fermion determinant in powers of
$\Nf$.  Unlike the 't~Hooft expansion case, at present,
\begin{equation}
 \bigl\langle \underbrace{\Gamma \Gamma \cdots \Gamma }_{k \text{ times}}
 \bigr\rangle_{{\rm c;YM}} \sim \Nc^2  \Nf^{k} \;,
\end{equation}
which means that the expansion converges if $\Nf$ is small enough.  It
was estimated that at $T=0$ the worldline expansion converges for
$\Nf < 4\Nc/(\Nc-2)$~\cite{Armoni:2009jn}, when the theory is below
the conformal window.

The rest of the discussion is almost identical to the previous
discussion.  The main issue is that $\muB=\frac{1}{2}\Nc(\Nc-1)\mu$.
The free energy $F^{(1)}/T$ at $T=0$ is $\mu$-independent as long as
$\mu$ is below the onset, according to 
\begin{equation}
 F^{(1)}/T \;\sim\; -\sum_{\bar{\omega}=-\infty}^\infty \exp\bigl[
  \bar{\omega}(\muB - \tilde{\MB} )/T \bigr] \;.
\label{eq:baryon2}
\end{equation}
The question is then how large $\tilde{\MB}$ should be?  Because the
large-$\Nc$ orientifold theory is not quenched, the $\Nf$
expansion around the Yang-Mills theory at $\Nf=0$ may be contaminated
by the introduction of the corresponding ghost $\tilde{\phi}$ that is
required in order to take the average $\langle\cdots\rangle_{\rm YM}$.

Therefore, if we could perform a quantitative comparison of $F^{(1)}$
calculated with the fundamental fermions and with the anti-symmetric
fermions, and if $\Nc$ is large enough, we could in principle verify
our analysis.  For the orientifold large-$\Nc$ case the scaling
between $\tilde{\MB}$ and the corresponding pion mass $\tilde{\Mpi}$
would not be affected, while in 't~Hooft large-$\Nc$ case it would
show a deviation with increasing $\Nc$, which signals weakened
ghosts.

\section{Discussions and Conclusions}
\label{sec:Discussion}

In this paper we discussed QCD with chemical potential in the
framework of 't~Hooft's $1/\Nc$ and Veneziano's $\Nf/\Nc$ expansions
(together with a brief discussion on the ``orientifold'' expansion).
Starting with the QCD partition function, we showed that the free
energy at $T=0$ is $\mu$-independent in the regime of small $\mu$
where the density of pions or baryons vanishes.

We showed explicitly that at any given order in $\Nf/\Nc$ the
fermionic determinant can be expanded in windings along the temporal
(or thermal) direction.  This expansion converges only for small
values of the chemical potential, the breakdown of which indicates the
onset of finite density.

In particular, at $\calO(\Nf\Nc)$ of the free energy, the isospin and
the quark chemical potentials are equivalent to each other.  The
density onset is associated with a baryonic configuration of $\Nc$
windings, which leads to a baryonic mass scale $\MB$.  A crucial
ingredient in our analysis is the role of the center symmetry, i.e.\
$\mathrm{Z}_{\Nc}$.  We used the fact that in the confining vacuum
center symmetry is unbroken.  Therefore, while the Dirac determinant
may depend on the chemical potential, at $T=0$, the only non-zero
contributions come from the zero (modulo $\Nc$) winding sector and the
rest vanishes due to $\mathrm{Z}_{\Nc}$ phase fluctuations.

At $\calO(\Nf^2)$ of the free energy, on the other hand, the isospin
and the quark chemical potentials are no longer equivalent.  While no
new dependence on the quark chemical potential appears at this order,
for the isospin chemical potential we encounter a mesonic
configuration of a quark and an anti-quark.  In that case the physical
picture is as follows:  the isospin density onset is characterized by
such mesonic configurations, the mass from which is $\Mpi$.

An important part of our analysis consists of the estimate of
$\MB /\Nc$ vs.\ $\Mpi/2$.  With Veneziano's expansion we need to
evaluate the operator expectation values with the pure Yang-Mills
vacuum, and when the finite-density operator is non-Hermitian, the
inevitable inclusion of conjugate quarks or ghosts makes the behavior
of $\MB$ unphysical.  We argue that we can extract the physical
information on $\MB$ thanks to 't~Hooft's large-$\Nc$ limit.  Based on
diagrammatic analysis, also, we provided an intuitive account for
$\MB/\Nc > \Mpi/2$ at large $\Nc$.  Thus, we conclude that there is
definitely a window between the onset of the pion condensation and the
finite baryon density in this particular limit.

Apart from the Silver Blaze puzzle, our expansion scheme is quite
unique on its own.  We can see an appreciable difference from the
standard large-$\Nc$ limit if we consider the free energy at finite
temperature $T$:  There is an exponentially small but non-zero
contribution to the free energy of a form,
$F/T \sim \exp[-(\MB-\muB)/T]$ in the baryon case and
$F/T \sim \exp[-(\mpi-2\muI)/T]$ in the isospin case.  Therefore, as
long as $T$ is not strictly zero, the free energy is \textit{not}
completely $T$-independent, as one might have naively expected from
the large-$\Nc$ Eguchi-Kawai reduction~\cite{Eguchi:1982nm}.  This is
because we first expanded the Dirac operator to identify the terms of
$\calO(\Nc\Nf)$ and $\calO(\Nf^2)$ and these sub-leading and
sub-sub-leading terms contain the baryonic and the mesonic
excitations, respectively, as ``valence'' degrees of freedom, though
virtual excitations are prohibited in the large-$\Nc$ limit.  It would
be worth revisiting Veneziano's limit to count the physical degrees of
freedom not only in the baryon sector~\cite{Hidaka:2008yy} but in the
meson sector also.

Future applications of our formulation would include concrete
implementation in the lattice-QCD simulation;  the investigation of
the Silver Blaze behavior in the canonical ensemble can clearly
separate two sectors sensitive to the isospin density and to the quark
density.  Then, one can test whether the onset is really given by the
physical pion mass and the physical baryon mass.  Once this is
confirmed, even without Veneziano's expansion and without potential
complication from ghosts, one may be able to justify our understanding
of the Silver Blaze behavior even for the case with finite $\Nc$.  On
the analytical level which would supplement the numerical efforts,
perhaps, the hopping-parameter expansion (which shares a similarity
with the large-$\Nc$ expansion), together with the re-summation of
higher-winding configurations~\cite{Green:1983sd}, will yield a useful
exemplification to diagnose the Silver Blaze puzzle, hopefully in a
consistent way with the scenario presented in this work.

\acknowledgments
  A.~A.\ wishes to thank the Kavli-IPMU for a warm and kind
  hospitality and to Gert~Aarts and Simon~Hands for discussions.
  The authors are grateful to Yoshimasa~Hidaka for intensive
  discussions. This work was supported by JSPS KAKENHI Grant No.~24740169.

\end{document}